\documentclass[twocolumn,showpacs,prl,superscriptaddress]{revtex4}
\usepackage{amsfonts}
\usepackage{amsmath}
\usepackage{amssymb}
\usepackage{amsthm}
\usepackage{graphicx}%
\pdfoutput=1
\newtheorem*{theorem}{Theorem}
\newcommand{\bra}[1]{\left\langle{#1}\right\vert}
\newcommand{\ket}[1]{\left\vert{#1}\right\rangle}
\newcommand{\braket}[2]{\left\langle{#1}\vert{#2}\right\rangle}
\begin{document}
\title{Localized closed timelike curves can perfectly distinguish quantum states}
\author{Todd A. Brun}
\affiliation{Communication Sciences Institute, Department of Electrical Engineering,
University of Southern California, Los Angeles, CA 90089, USA}
\author{Jim Harrington}
\affiliation{Applied Modern Physics (P-21), MS D454, Los Alamos National Laboratory, Los Alamos,
NM 87545, USA}
\author{Mark M. Wilde}
\affiliation{Communication Sciences Institute, Department of Electrical Engineering,
University of Southern California, Los Angeles, CA 90089, USA}
\affiliation{Centre for Quantum Technologies, National University of Singapore, 3 Science
Drive 2, Singapore 117543}
\date{\today}

\begin{abstract}
We show that qubits traveling along closed timelike curves are a resource that
a party can exploit to distinguish perfectly any set of quantum states. As a
result, an adversary with access to closed timelike curves can break any
prepare-and-measure quantum key distribution protocol. Our result also implies
that a party with access to closed timelike curves can violate the Holevo bound.
\end{abstract}

\maketitle


\textit{Introduction}---The theory of general relativity points to the possible 
existence of closed timelike curves (CTCs) \cite{PhysRevLett.61.1446, gott91}. 
The \textit{grandfather paradox} is one criticism raised to their existence, but 
Deutsch resolved this paradox by presenting a method for finding self-consistent 
solutions of CTC interactions \cite{PhysRevD.44.3197}.  

Recently, several quantum information researchers have assumed that CTCs exist 
and have examined the consequences of this assumption for \textit{computation}
\cite{fpl2003brun,PhysRevA.70.032309,arx2008aaronson}.  Brun showed that a 
classical treatment (assuming a lack of contradictions) allows NP-hard problems to 
be computed with a polynomial number of gates \cite{fpl2003brun}. Bacon followed 
with a purely quantum treatment that demonstrates the same reduction of NP-hard 
problems to P, along with a sketch of how to perform this reduction in a fault-tolerant 
manner \cite{PhysRevA.70.032309}. Aaronson and Watrous have recently established 
that either classical or quantum computers interacting with closed timelike curves
can compute any function in PSPACE in polynomial time \cite{arx2008aaronson}.

In this Letter, we show how a party with access to CTCs, or a \textquotedblleft 
CTC-assisted\textquotedblright\ party, can perfectly distinguish among a set of 
non-orthogonal quantum states. The result has implications for fundamental
protocols in quantum \textit{communication} because a simple corollary is 
that a CTC-assisted party can break any prepare-and-measure quantum key 
distribution protocol \cite{b92,bb84,sarg04}. 
(The security of such a scheme relies on the information-disturbance tradeoff for 
identifying quantum states.) Furthermore, the capacity for quantum systems to
carry classical information becomes unbounded.  

Our work here raises fundamental questions concerning the nature of a physical
world in which closed timelike curves exist because it challenges the postulate of 
quantum mechanics that non-orthogonal states cannot be perfectly distinguished. 
A full theory of quantum gravity would have to resolve this apparent contradiction 
between the implication of CTCs and the laws of quantum mechanics.  Note that any
alternative source of nonlinearity would raise similar questions.

We structure this Letter as follows. First, we give some background on
Deutsch's formalism regarding CTCs in quantum information theory
\cite{PhysRevD.44.3197}. We then show how to distinguish the non-orthogonal
states $\left\vert {0}\right\rangle $ and $\left\vert {-}\right\rangle $ where
$\left\vert {-}\right\rangle \equiv(\left\vert {0}\right\rangle -\left\vert
{1}\right\rangle )/\sqrt{2}$ and follow by showing how to distinguish the
\textquotedblleft BB84\textquotedblright\ states $\left\vert {0}\right\rangle
$, $\left\vert {1}\right\rangle $, $\left\vert {+}\right\rangle $, and
$\left\vert {-}\right\rangle $ where $\left\vert {+}\right\rangle
\equiv(\left\vert {0}\right\rangle +\left\vert {1}\right\rangle )/\sqrt{2}$.
Our main theorem then shows that a CTC-assisted party can 
perfectly distinguish among an arbitrary set of states. We end by discussing 
how a CTC-assisted party can break Holevo's bound \cite{holevo}.

\textit{Background}---Qubits traveling around closed timelike curves (CTC
qubits) may give rise to highly nonintuitive behavior, but Deutsch showed how
to avoid certain paradoxes by imposing a self-consistency condition
\cite{PhysRevD.44.3197}. This self-consistency condition requires that the input
density matrix of a CTC quantum system match its output density matrix
following its interaction with another system:
\begin{equation}
\rho_{\text{CTC}}={\text{Tr}}_{\text{sys}}\{V\left(  \left\vert {\psi
}\right\rangle \left\langle {\psi}\right\vert \otimes\rho_{\text{CTC}}\right)
V^{\dagger}\},
\label{rho_CTC}
\end{equation}
where $\left\vert {\psi}\right\rangle$ is the input state of the chronology-respecting 
system, the matrix $\rho_{\text{CTC}}$ is the initial density matrix of the CTC 
quantum system before the two systems interact, and $V$ is the interaction unitary. 
The expression on the right hand side of (\ref{rho_CTC}) is the partial density 
matrix of the CTC system after the interaction.
The output state of the chronology-respecting system is then
\begin{equation}
\rho_{\text{out}}={\text{Tr}}_{\text{CTC}}\{V\left(  \left\vert {\psi
}\right\rangle \left\langle {\psi}\right\vert \otimes\rho_{\text{CTC}}\right)
V^{\dagger}\}.
\label{rho_out}
\end{equation}
The output state is in general a nonlinear function of the input state 
$\left\vert {\psi}\right\rangle$, because $\rho_{\text{out}}$ depends on both 
$\left\vert {\psi}\right\rangle $ and $\rho_{\text{CTC}}$, and $\rho_{\text{CTC}}$ 
also depends on $\left\vert {\psi}\right\rangle $. It is this nonlinearity that 
enables us to transcend the usual limitations of quantum mechanics.

Deutsch showed in Ref.~\cite{PhysRevD.44.3197} that there always exists a self-consistent 
solution to Eq.~(\ref{rho_CTC}), but it does not necessarily have to be unique.  In the 
examples and main theorem of this Letter, we construct an interaction and measurement 
scheme to distinguish perfectly any set of non-orthogonal states.  To achieve this result, we 
engineer the density matrix of the CTC system to be unique as well as self-consistent.

\textit{Distinguishing two non-orthogonal states}---We first show how to
distinguish the non-orthogonal states $\left\vert {0}\right\rangle $ and
$\left\vert {-}\right\rangle $ without uncertainty or error. Let $\left\vert
{\psi}\right\rangle ^{A}$ denote the unknown initial state ($\left\vert
{0}\right\rangle $ or $\left\vert {-}\right\rangle $) that lives on a system
$A$. Suppose that we have access to one CTC qubit for a length of time and let
$B$ denote its corresponding system. The desired interaction is as follows:

\begin{enumerate}
\item ${\text{Swap }}$systems $A$ and $B$.
\item Perform a controlled-Hadamard with system $A$ as the control and system
$B$ as the target.
\item Measure system $A$ in the computational basis.
\end{enumerate}

\begin{figure}[ptb]
\includegraphics[width=3in]{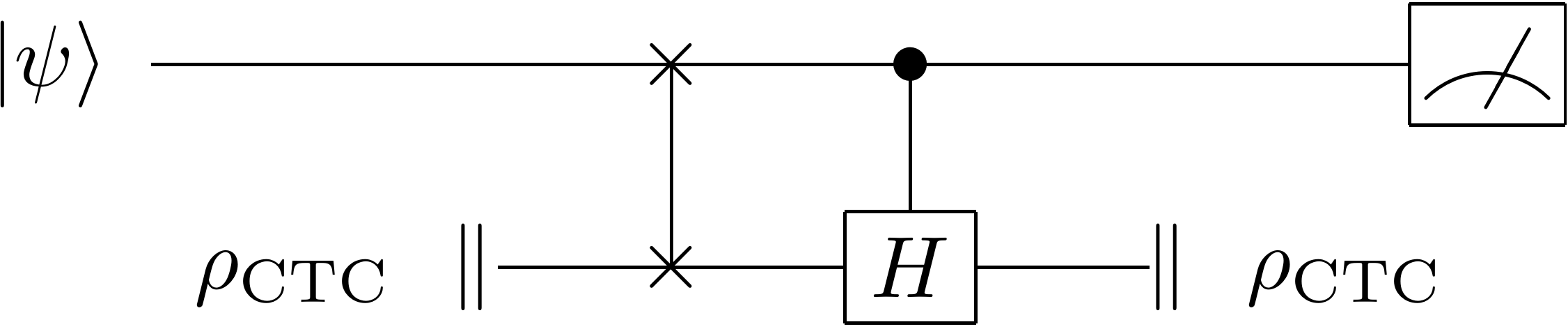}\caption{The above circuit can
perfectly distinguish the non-orthogonal states $\left\vert {0}\right\rangle $
and $\left\vert {-}\right\rangle $. The first qubit in state $\left\vert
{\psi}\right\rangle $\ is the unknown qubit ($\left\vert 0\right\rangle $ or
$\left\vert -\right\rangle $) and the second qubit with density matrix
$\rho_{\text{CTC}}$ travels along a closed timelike curve. The double vertical
bars on the bottom left and right indicate the past and future mouths of the
wormhole for the CTC.}%
\label{fig:b92}%
\end{figure}

System $B$ ``disappears'' after some time because it travels along a closed
timelike curve and enters the future mouth of its wormhole.  The measurement 
of system $A$ occurs after this point.  A measurement result of zero reveals 
that $\left\vert {\psi}\right\rangle =\left\vert {0}\right\rangle $, and a measurement 
result of one reveals that $\left\vert {\psi}\right\rangle=\left\vert {-}\right\rangle$. 
Fig.~\ref{fig:b92} depicts the quantum circuit for this procedure. 

Let us describe the operation of the circuit in Fig.~\ref{fig:b92} by tracing 
backward through it. First suppose that the final state of the chronology-respecting 
qubit is $\left\vert {0}\right\rangle \left\langle {0}\right\vert $. The circuit is then 
simply a {\textrm{SWAP}} gate because the Hadamard does not act on the CTC 
qubit. Therefore, self-consistency of the initial and final state of the CTC qubit 
implies that $\rho_{\text{CTC}}=\left\vert {\psi}\right\rangle \left\langle {\psi}\right\vert 
=\left\vert {0}\right\rangle\left\langle {0}\right\vert$ because the two qubits are 
invariant under the ${\text{SWAP}}$ operation.

Alternatively, suppose the final state of the chronology-respecting qubit is
$\left\vert {1}\right\rangle \left\langle {1}\right\vert $. Then the
controlled-Hadamard reduces to application of the Hadamard gate on the CTC
qubit. The input state to the Hadamard gate is $\left\vert {\psi}\right\rangle
\left\langle {\psi}\right\vert $ (because of the ${\text{SWAP}}$), and the
output state is $\rho_{\text{CTC}}=\left\vert {1}\right\rangle \left\langle
{1}\right\vert $ (again, because of the ${\text{SWAP}}$). This action occurs
when $\left\vert {\psi}\right\rangle \left\langle {\psi}\right\vert
=\left\vert {-}\right\rangle \left\langle {-}\right\vert $.

It only remains to show that these self-consistent solutions for
$\rho_{\text{CTC}}$ are unique. Let
\[
\rho_{\text{CTC}}=\alpha\left\vert {0}\right\rangle \left\langle
{0}\right\vert +\beta\left\vert {0}\right\rangle \left\langle {1}\right\vert
+\gamma\left\vert {1}\right\rangle \left\langle {0}\right\vert +\delta
\left\vert {1}\right\rangle \left\langle {1}\right\vert .
\]
For $\rho_{\text{CTC}}$ to be a density matrix, it must be Hermitian, positive 
semi-definite, and have trace 1; these conditions imply that $\alpha,\delta$ must 
be non-negative reals such that $\alpha+\delta=1$, that $\gamma=\beta^{\ast}$,
and that $|\beta|^{2}\le\alpha\delta$. Suppose $\left\vert {\psi}\right\rangle
\left\langle {\psi}\right\vert =\left\vert {0}\right\rangle \left\langle{0}\right\vert $. 
Then $\delta=0$ and $\alpha=1$ because self-consistency requires that 
$\alpha=\alpha+\delta/2$. 
Thus $\rho_{\text{CTC}}=\left\vert{0}\right\rangle \left\langle {0}\right\vert $ is the 
only solution. Now suppose
 $\left\vert {\psi}\right\rangle \left\langle {\psi}\right\vert
=\left\vert {-}\right\rangle \left\langle {-}\right\vert $. 
Then $\alpha=0$ and $\delta=1$ because self-consistency requires that 
$\delta=\delta+\alpha/2$. Thus 
$\rho_{\text{CTC}}=\left\vert {1}\right\rangle \left\langle{1}\right\vert $ is the only solution.

Straightforward modifications to the unitaries in Fig.~\ref{fig:b92} can be
introduced to distinguish between any two non-orthogonal states. This scheme 
then breaks the security of the B92 quantum key distribution protocol \cite{b92}. 
Even with no loss on the quantum channel, a CTC-assisted adversary can learn 
the identity of every signal that Alice transmits and then prepare and transmit the 
same state to Bob. The adversary gains full information without producing any 
disturbance.

\begin{figure}[ptb]
\includegraphics[width=3in]{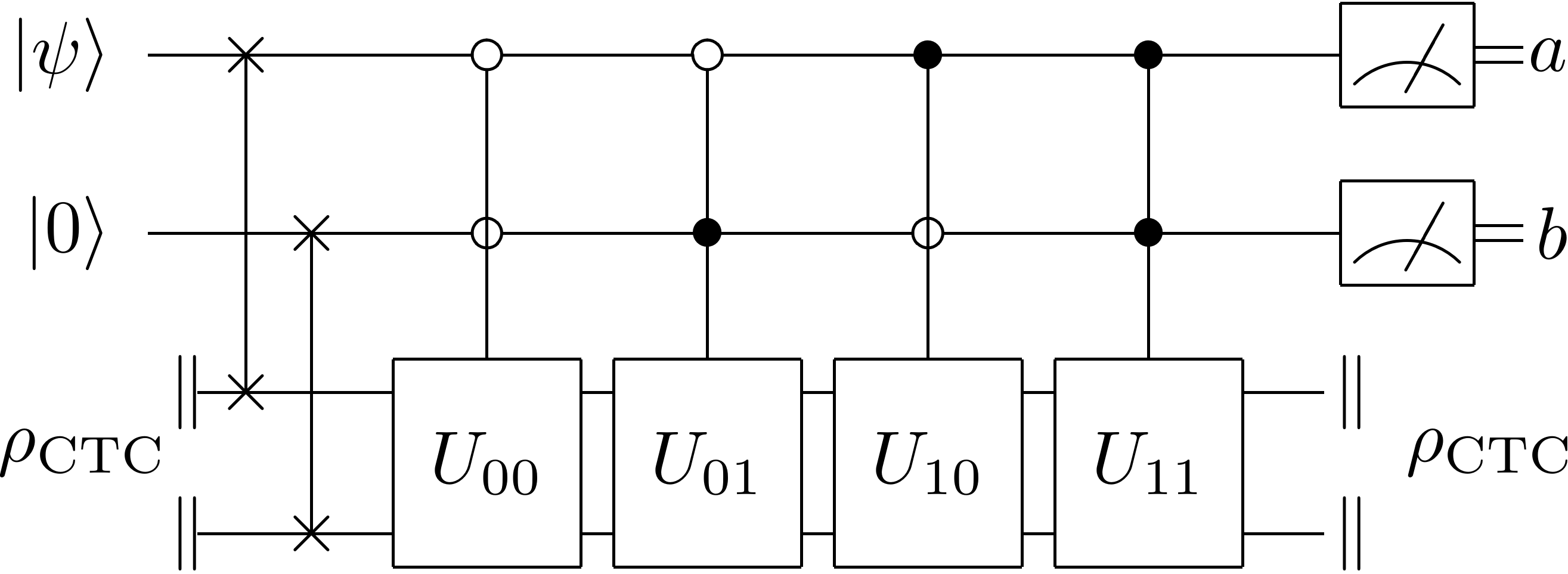}\caption{The above circuit can
perfectly distinguish the BB84 states $\left\vert {0}\right\rangle $,
$\left\vert {1}\right\rangle $, $\left\vert {+}\right\rangle $, and
$\left\vert {-}\right\rangle $. The circuit uses the standard quantum circuit
notation from Ref.~\cite{book2000mikeandike} and we define the unitaries
$U_{00}$, $U_{01}$, $U_{10}$, and $U_{11}$ in (\ref{eq:break-bb84}).}%
\label{fig:bb84}%
\end{figure}

\textit{Distinguishing the BB84 states}---We next consider how to distinguish
the four BB84 states $\{\left\vert {0}\right\rangle ,\left\vert {1}%
\right\rangle ,\left\vert {+}\right\rangle ,\left\vert {-}\right\rangle \}$.
Our scheme first appends an ancillary state $\left\vert {0}\right\rangle $ to
the unknown state $\left\vert {\psi}\right\rangle $ (one of the four BB84
states) and then uses two CTC qubits to effect the following map:
\begin{align*}
\left\vert {00}\right\rangle \rightarrow\left\vert {00}\right\rangle ,\ \ \  &
\ \ \ \left\vert {+0}\right\rangle \rightarrow\left\vert {10}\right\rangle ,\\
\left\vert {10}\right\rangle \rightarrow\left\vert {01}\right\rangle ,\ \ \  &
\ \ \ \left\vert {-0}\right\rangle \rightarrow\left\vert {11}\right\rangle .
\end{align*}
That is, by measuring the output of the chronology-respecting qubits in the
computational basis, the result $a=0$ reveals that the unknown state
$\left\vert {\psi}\right\rangle $ is a $Z$-eigenstate with eigenvalue
$(-1)^{b}$, and $a=1$ reveals that $\left\vert {\psi}\right\rangle $ is an
$X$-eigenstate with eigenvalue $(-1)^{b}$. We claim that the circuit in
Fig.~\ref{fig:bb84} implements such a mapping, where we define the
unitaries $U_{00}$, $U_{01}$, $U_{10}$, and $U_{11}$ as follows:
\begin{align}
U_{00}  &  \equiv{\text{SWAP,}}\nonumber\\
U_{01}  &  \equiv X\otimes X,\nonumber\\
U_{10}  &  \equiv(X\otimes I)\circ(H\otimes I),\nonumber\\
U_{11}  &  \equiv(X\otimes H)\circ({\text{SWAP}}).\label{eq:break-bb84}%
\end{align}

The circuit in Fig.~\ref{fig:bb84} consists of two SWAPs between the
chronology-respecting qubits and the CTC qubits, followed by four controlled
unitaries, such that a distinct unitary acts on the CTC qubits for each output
state $\left\vert {ab}\right\rangle $. For each input state, the desired
output of the chronology-respecting qubits corresponds to a self-consistent
solution for the CTC qubits. The argument that the solution is unique proceeds
as before: we consider a general density matrix for $\rho_{\text{CTC}}$, and we 
then show that all but one of the diagonal elements in the computational basis must
be zero. This result implies that $\rho_{\text{CTC}}$ is pure and equal to a
computational basis state. 

As in the previous section, the circuit in Fig.~\ref{fig:bb84} renders insecure any 
quantum key distribution protocol using these states \cite{bb84,sarg04}.
An adversary can learn the basis and bit values of each signal state (and then prepare 
an identical state) without introducing any loss or disturbance in the quantum transmission.

\textit{General state distinguishability}---We now present our main theorem and proof,
that constructively demonstrates how to use a CTC system to distinguish
perfectly an arbitrary number of distinct quantum states.

\begin{theorem}
Suppose there is a set $\left\{  \left\vert \psi_{j}\right\rangle \right\}
_{j=0}^{N-1}$\ of $N$ distinct states in a space of dimension $N$. Suppose we
have access to an $N$-dimensional CTC system in a closed loop. Then we can
implement the following map:
\[
\forall j\ \ \ \ \left\vert \psi_{j}\right\rangle \rightarrow\left\vert
j\right\rangle
\]
where the states $\left\vert j\right\rangle $ are a standard orthonormal basis
for the $N$-dimensional space.
\end{theorem}

\begin{proof}
We want to demonstrate a mapping of $\ket{\psi_j} \rightarrow\ket{j}$ for $0
\leq j \leq N-1$, where $\{\ket{j}\}$ forms a standard orthonormal basis for
the input space. We utilize a closed timelike curve (CTC) containing an
$N$-dimensional system in a closed loop.
We prepare the input system in one of the states $\ket{\psi_j}$. We then let
it interact with the CTC system via a unitary transformation $V$. The output
state will be $\ket{j}$. We choose $V$ as follows:
\begin{enumerate}
\item First, swap the input system with the CTC system.
\item Next, apply the following controlled unitary from the system to the
CTC:
\[
\sum_{k=0}^{N-1}\ket{k}\bra{k}\otimes U_{k},
\]
where the $\{U_{k}\}$ are a set of $N$ unitary transformations acting just on
the CTC system.
\end{enumerate}
Let the input state of the chronology-respecting system by $\ket{\psi_j}$.
Before the interaction, the CTC system is in the state $\rho_{\text{CTC}}$,
which must satisfy the self-consistency condition Eq.~(\ref{rho_CTC}) for
$\ket\psi = \ket{\psi_j}$.  The state of the output system will be given by
Eq.~(\ref{rho_out}).  We first show how to satisfy self-consistency. 
If we choose each $U_{k}$ such that
\begin{equation}
U_{k}\ket{\psi_k}=\ket{k},
\label{consistency1}
\end{equation}
then the solution $\rho_{\text{CTC}}=\ket{k}\bra{k}$ satisfies the
self-consistency condition and gives the desired output state.
However, this is not enough by itself for the construction to work. We also
need $\rho_{\text{CTC}}$ to be unique. (More precisely, $\rho_{\text{out}}$
needs to be unique. But uniqueness of $\rho_{\text{CTC}}$ is a
sufficient condition for that.)
We now show how to engineer the state of the CTC\ system and the output system
to be unique. Suppose that the $\{U_{k}\}$ satisfy the condition above.
Consider a general state for $\rho_{\text{CTC}}$:
\[
\rho_{\text{CTC}}=\sum_{m,n}\rho_{mn}\ket{m}\bra{n}.
\]
Plugging this expression into the self-consistency equation (\ref{rho_CTC}) for
$\rho_{\text{CTC}}$ with input state $\ket{\psi_j}$ and a unitary $V$ of the above
form, the matrix elements $\rho_{mn}$ must satisfy
\begin{equation}
\rho_{mn}=\sum_{k}\rho_{kk}\bra{m}U_{k}\ket{\psi_j}\bra{\psi_j}U_{k}^{\dagger}\ket{n}.
\label{consistency2}
\end{equation}
We want to choose the unitaries $\{U_{k}\}$ such that the unique solution to
Eq.~(\ref{consistency2}) is $\rho_{jj} = 1$, and all other elements of
$\rho_{\text{CTC}}$ are zero.  Let us focus on the $j$th diagonal element.
Since $U_{j}\ket{\psi_j}=\ket{j}$, we get
\begin{equation}
\rho_{jj}=\rho_{jj}+\sum_{k\neq j}\rho_{kk}\left\vert \bra{j}U_{k}
\ket{\psi_j}\right\vert ^{2}.
\label{consistency3}
\end{equation}
For any $k$ such that $\bra{j}U_{k}\ket{\psi_j}\neq0$, the above equation
implies $\rho_{kk}=0$. If $\rho_{kk}=0$ for all $k\neq j$, this implies
that all off-diagonal terms are also zero, and therefore $\rho_{jj}=1$, which
is what we want. Therefore, a set of sufficient (but by no means necessary) 
conditions for a unique, self-consistent solution are as follows:
\begin{enumerate}
\item $U_{k} \ket{\psi_k} = \ket{k}$ for all $k$, and
\item $\bra{j} U_{k} \ket{\psi_j} \ne0$ for all $j$ and $k$.
\end{enumerate}
Next we construct a set of unitaries $\{U_{k}\}$ satisfying these two conditions.
Let $S=\{\ket{\psi_j}\}$ be the set of initial states. Choose a particular
$k$. We will construct two orthonormal bases $\ket{b_m}$ and $\ket{c_m}$ for
$m=1,\ldots,N$ such that
\[
U_{k}=\sum_{m}\ket{c_m}\bra{b_m}.
\]
This will automatically make $U_{k}$ unitary. We construct these bases in a
series of steps.
\vspace{.1in}
\newline
1. We need $U_{k}\ket{\psi_k}=\ket{k}$. So choose $\ket{b_1}=\ket{\psi_k}$
and $\ket{c_1}=\ket{k}$.  Let us  label the vector $\ket{\psi_k}$ as $\ket{\psi_{1,1}}$.
\vspace{.1in}
\newline
2. Pick another vector from the set $S$. Label this vector
$\ket{\psi_{2,1}}$. Perform a Gram-Schmidt orthogonalization 
with this vector to construct orthonormal basis vector $\ket{b_2}$:
\[
\ket{b_2}=\frac{1}{\mathcal{N}}\left(\ket{\psi_{2,1}}-\ket{b_1}\left\langle
{b_{1}}|{\psi_{2,1}}\right\rangle \right).
\]
3. Now find all the vectors in the set $S$ that are in the space spanned by
$\ket{b_1}$ and $\ket{b_2}$, including at least $\ket{\psi_{2,1}}$,
but  excluding $\ket{\psi_{1,1}}$. Suppose there are $m_{2}$ such vectors. Label these
vectors $\ket{\psi_{2,1}}, \ket{\psi_{2,2}}, \ldots,\ket{\psi_{2,m_2}}$.
Construct the basis vector $\ket{c_2}$:
\[
\ket{c_2}=\frac{1}{\sqrt{m_{2}}}\left(  \sum_{n=1}^{m_{2}}\ket{j_{2,n}}\right) ,
\]
where the labels $j_{2,n}$ stand for the indices of the vectors $\ket{\psi_{2,n}}$
in the set. Note that $\ket{c_2}$ is also orthogonal to $\ket{c_1}$.
\vspace{.1in}
\newline
4. We now iterate this procedure. Suppose we have constructed $t$ basis
vectors $\ket{b_1},\ldots,\ket{b_t}$ and $\ket{c_1},\ldots,\ket{c_t}$. We
construct $\ket{b_{t+1}}$ and $\ket{c_{t+1}}$ as follows. Pick a state from $S$
that has not yet been used. Label this state $\ket{\psi_{t+1,1}}$. Perform a
Gram-Schmidt orthogonalization using this state and the already constructed vectors
$\ket{b_1},\ldots,\ket{b_t}$ to make the orthonormal basis vector $\ket{b_{t+1}}$:
\[
\ket{b_{t+1}}=\frac{1}{\mathcal{N}}\left(  \ket{\psi_{t+1,1}}-\sum
_{n=1}^{t}\ket{b_n}\braket{b_n}{\psi_{t+1,1}}\right)  .
\]
5. Take all the vectors from $S$ that have not yet been used and that are
contained in the subspace spanned by $\ket{b_1},\ldots,\ket{b_{t+1}}$. Suppose
there are $m_{t+1}$ of them. Label these vectors
$\ket{\psi_{t+1,1}},\ldots,\ket{\psi_{t+1,m_{t+1}}}$. Now construct
the new basis vector $\ket{c_{t+1}}$:
\[
\ket{c_{t+1}}=\frac{1}{\sqrt{m_{t+1}}}\left(  \sum_{n=1}^{m_{t+1}}\ket{j_{t+1,n}}\right)  .
\]
6. Repeat steps 4 and 5 until all the vectors in the set $S$ have been
used. If this has not yet produced a complete basis, choose any sets of
orthonormal vectors to complete $\{\ket{b_m}\}$ and $\{\ket{c_m}\}$.
\vspace{.1in}
\newline
7. Now repeat this entire construction for every $U_{k}$. From step 1 we
get condition 1: $U_{k}\ket{\psi_k}=\ket{k}$. From the way we construct the
$\ket{c_m}$ (in steps 3 and 5), we see that $\bra{j}U_{k}\ket{\psi_j}\neq0$
for all $j$ and $k$, so both self-consistency and uniqueness are assured.
\end{proof}

\textit{Implications for the Holevo bound}---As a final note, we point out
that a CTC-assisted party can violate the Holevo bound \cite{holevo}. 
Suppose that Alice chooses to send one of the four states 
$\{\left\vert {0}\right\rangle, \left\vert {1}\right\rangle, 
\left\vert {+}\right\rangle, \left\vert {-}\right\rangle \}$ 
to Bob over a noiseless quantum channel. A CTC-assisted Bob can 
employ the method in the previous section to distinguish Alice's state 
perfectly and can then access two classical bits of information.  This 
ability to access two classical bits violates the Holevo bound of one 
classical bit per qubit. Indeed, using a set of $2^{n}$ non-orthogonal 
states would allow Alice to send $n$ classical bits via a single 
noiseless qubit, if Bob uses the above measurement procedure.

\textit{Conclusion}---We have shown how to exploit closed timelike curves to 
distinguish non-orthogonal states. Two direct implications are that one could 
break any prepare-and-measure quantum key distribution protocol as well as 
violate the Holevo bound.   If CTC qubits are treated as a free resource, then 
the achievable classical communication rate with a single noiseless quantum 
transmission is unbounded.  We conjecture that the addition of any nonlinearity 
to quantum mechanics, such as that considered in Ref.~\cite{abramslloyd98}, 
could be exploited similarly.

There are at least three ways to consider the implications of the results in this Letter.  
First, note that even if our universe contains no stable wormholes, the existence of 
microscopic, short-lived closed timelike curves can still revolutionize information 
processing tasks if they persist long enough to engineer specific unitary interactions 
with qubits traveling their worldlines.  
Second, while issues such as the grandfather paradox are resolved by Deutsch's 
formalism for stochastic and quantum bits traveling along closed timelike curves 
\cite{PhysRevD.44.3197}, the eroding of a finite capacity for classical 
communication with a qubit is a strong information theoretic argument casting
doubt on the allowed existence of CTCs (similar in vein to the quantum 
communication complexity argument in Ref.~\cite{trivialcommunicationcomplexity}).
A third tack is to consider whether Deutsch's fixed point solution for resolving CTC 
paradoxes is itself somehow flawed.  If the formalism is invalidated, then 
computational complexity results such as $\text{P}_{\text{CTC}} = \text{PSPACE}$ 
\cite{arx2008aaronson} should be reexamined.  Any theory of quantum gravity will 
need to reconcile this intersection of quantum information theory and general relativity.  

Finally, it should be interesting to study the effect of noise on the physical processes 
outlined in this Letter. For instance, how stable are these maps to perturbations in the 
input states?  Recent work utilizing the Heisenberg picture may be a useful approach 
\cite{ralph07}.  We conjecture that a CTC-assisted party can construct a universal cloner 
with fidelity approaching one, at the cost of increasing the available dimensions in 
ancillary and CTC resources. One area of future work could be to optimize this fidelity 
given CTC resources of fixed dimension.

\begin{acknowledgments}
We thank Dave Bacon, Steve Flammia, Charlie Bennett, Tim Ralph and Jonathan 
Oppenheim for helpful discussions.  MMW\ acknowledges support from NSF Grant 
0545845, from the National Research Foundation \& Ministry of Education, Singapore, 
and thanks Martin R\"{o}tteler and  NEC\ Laboratories America for hosting him as a visitor. 
TAB received support from NSF\ Grant No.~CCF-0448658.
\end{acknowledgments}

\bibliographystyle{apsrev}
\bibliography{PRL_CTC_QKD}

\end{document}